\documentclass[aps,twocolumn,footinbib,showpacs,floatfix]{revtex4}
\usepackage{amssymb}
\usepackage[pdftex]{graphicx}	
	\DeclareGraphicsExtensions{.pdf,.png,.jpg}
\usepackage[applemac]{inputenc}
\usepackage{amsmath,amssymb,amsthm}
\usepackage{mathbbol,bbm}
\usepackage{graphicx,float}
\usepackage[final]{showkeys}
\usepackage{bm,dsfont}

\def\Tr{ {\rm{Tr }}}

 \def\ee{\mathord{\rm e}}
 
 \def\ii{\mathord{\rm i}}

\renewcommand{\ii}{{\rm i}}
\renewcommand{\ee}{{\rm e}}

\begin{document}
\title{Thermal Instability of Protected End States
in a One-Dimensional Topological Insulator}
\author{O. Viyuela, A. Rivas and M.A. Martin-Delgado}
\affiliation{Departamento de F\'{\i}sica Te\'orica I, Universidad Complutense, 28040 Madrid, Spain}

\vspace{-3.5cm}

%\date{\today}
\begin{abstract}
We have studied the dynamical thermal effects on the protected end states of a topological insulator (TI)
when it is considered as an open quantum system in interaction with a noisy environment at a certain temperature $T$.
As a result, we find that protected end states in a TI become unstable and decay with time.
Very remarkably, the interaction with the thermal environment (fermion-boson) respects chiral symmetry, which is
the symmetry responsible for the protection (robustness) of the end states in this TI
when it is isolated from the environment. Therefore, this mechanism makes end states unstable while preserving
their protecting symmetry. Our results have immediate practical implications in recently proposed simulations of TI using cold atoms in optical lattices. Accordingly, we have computed lifetimes of topological end states for
these physical implementations that are useful to make those experiments realistic.
\end{abstract}

\maketitle

%%%%
\section{Introduction}
%%%%%

The stability of topological phases of matter, also known as topological orders \cite{wenbook},
against thermal noise has provided several surprising results in the context of topological codes used
in topological quantum information \cite{AHF09,nussinov_ortiz_08}. However, very little is known about the behavior of a topological insulator (TI) subject to the disturbing thermal effect of its surrounding environment. This is of great relevance if we want
to address key questions such as the robustness of TIs to thermal noise, existence of thermalization processes, use of TIs as platforms for quantum computation, etc.
Topological insulators have emerged as a new type of quantum phase of matter \cite{rmp1,rmp2} that was predicted theoretically to exist \cite{haldane_88,kane_mele_05,bernevig_zhang_06,fu_kane_07,fu_kane_mele_07,moore_balents_07,qi_hughes_zhang_08,roy_09} and
has been discovered experimentally \cite{TI_exp1, TI_exp3D, TI_exp2}. Exploring the possible features and uses of TIs has become a very active interdisciplinary field.
For this,  knowledge about their stability under nonequilibrium thermal dynamics is crucial in assessing the feasibility of proposals in
quantum computation, spintronics, etc.

In this work we present a first-principle calculation to test several thermal effects on a one dimensional (1D) TI out of equilibrium.
In order to achieve these goals, we need first to specify two choices: the type of TI and the type of thermal baths.
As for TI, we work with the Creutz Ladder (CL) which is a paradigmatic example of a quasi-one-dimensional fermion system
that exhibits the fundamental properties of TIs. Namely, localized states in the bulk of system and end states in the form of zero-energy modes at the boundary \cite{Creutz, Berm1}.
It is a crucial remark that the presence of these end states is independent of whether the system size is finite or infinite. They constitute a clear signature of a TI in the case of the CL.
Although the first experimental realizations of TIs are in 2D and 3D, the case of 1D TIs also appears in the so-called `Periodic Table' of TIs \cite{ptable1, ptable2,comment1}. Moreover, there are recent proposals to realize TIs in 1D optical lattices \cite{1D_TI_OL_12, 1D_TI_QC_11}, and in particular the CL \cite{CL_OL_12}.

%%%%%%%%%%%%%%%%%%%%%%%%%%
%%%%%%%%%%%%%%%%%%%%%%%%%%
\begin{figure}[t]
	\includegraphics[width=\columnwidth]{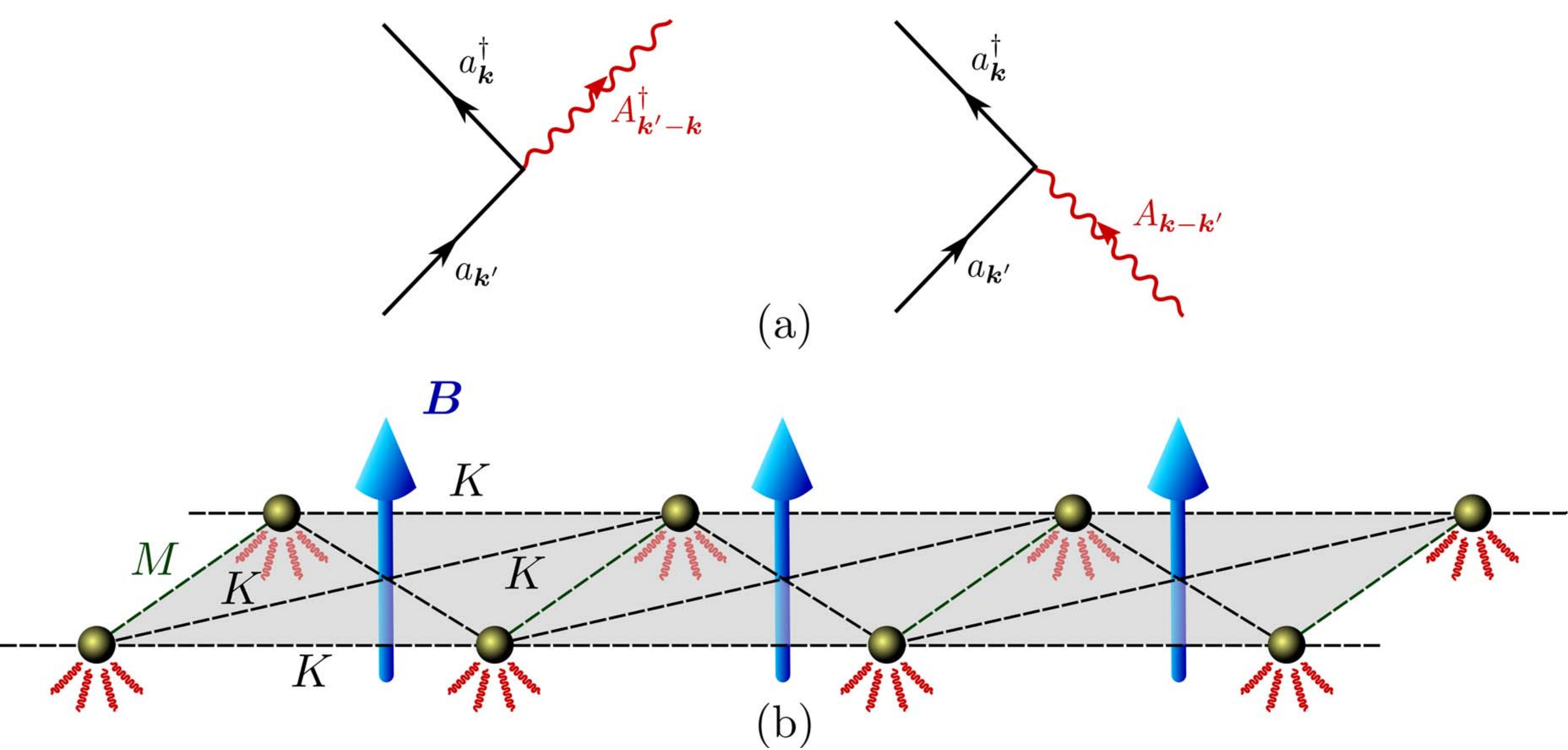}
	\caption{
 The TI system is pictured in (b) as a ladder of hopping spinless fermions with amplitudes $K$ (horizontal and diagonal) and $M$ (vertical).
 The fermions are coupled to a magnetic gauge field on the lattice that is perpendicular to the plaquettes. This is a  lattice gauge theory known as the Creutz Ladder
 (CL) \cite{Creutz,Berm1}. The wavy lines at each sites indicate interaction with thermal bosonic baths. In (a),  the interaction vertex of fermions with the bosonic bath
 is shown.}
	\label{fig:creutz_ladder}
\end{figure}
%%%%%%%%%%%%%%%%%%%%%%%%%%
%%%%%%%%%%%%%%%%%%%%%%%%%%

As for the environmental quantum noise, we model it in the form of local bosonic thermal baths (see Fig. \ref{fig:creutz_ladder}). This is rather
natural, but novel since usually the bath and system degrees of freedom
are taken to be of the same type. Here, we deal with a fermionic system but the bath is made up of bosons. The reason for this choice is inspired by the traditional electron-phonon interaction in crystal solids \cite{AM_76}.
The main difference is that our thermal baths are local in order to simplify their study. Moreover, this locality also fits into the traditional scheme of perturbing the
global properties of a topological order by means of local external noise, as it is natural in topological quantum information. If the CL is realized with optical lattices, then these local baths
can be though of as external photons. Therefore, the meaning of the bosonic baths will depend on the specific realization we choose.

\begin{figure}[h]
	\includegraphics[width=\columnwidth]{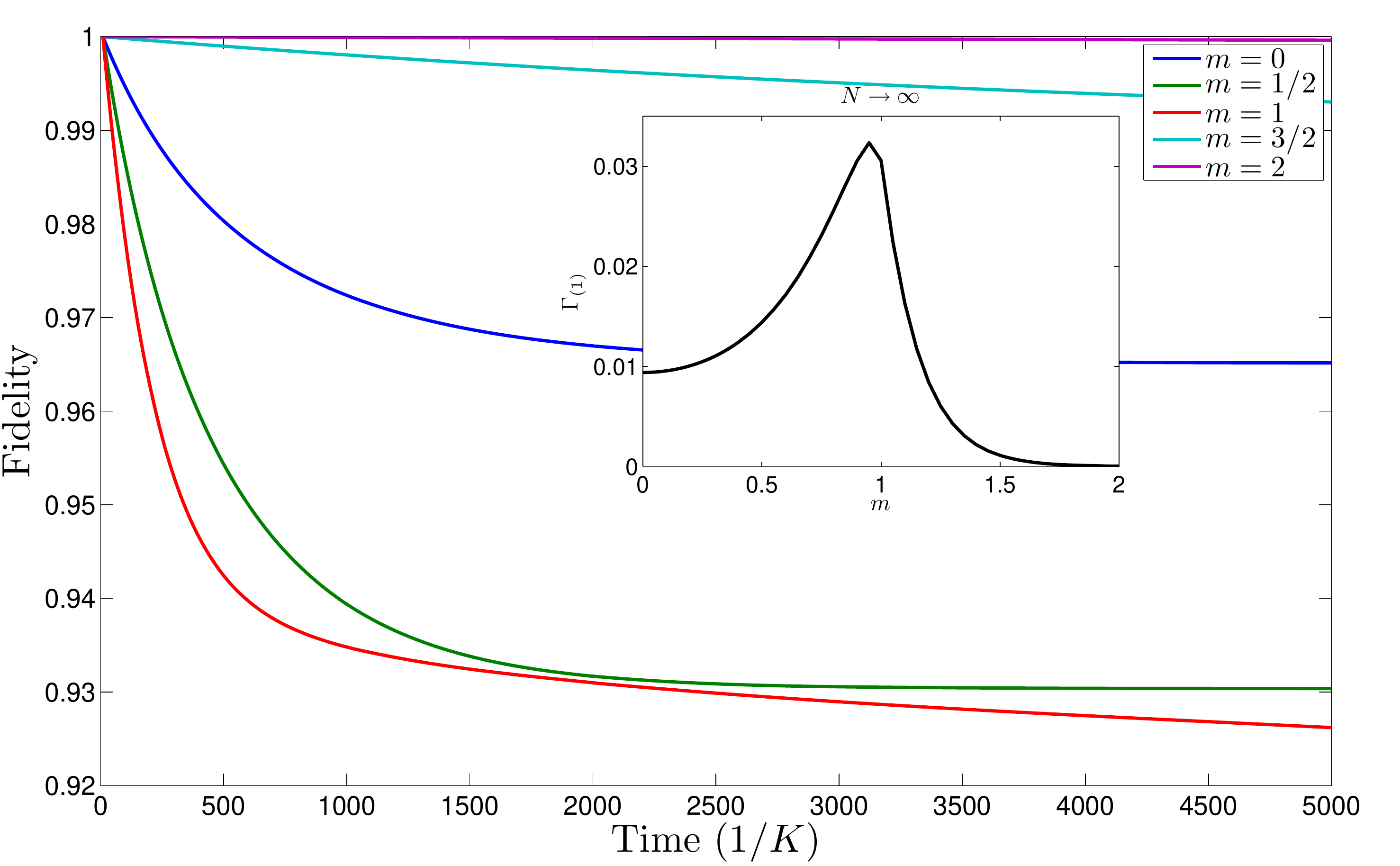}
	\caption{Fidelities for the state of the CL with the initial Fermi Sea. Here we have taken $\theta=\pi/2$, $T=1$ (in units of $K$, Eq. (2)) and $\omega_c=3/(2K)$ (of the order of the mean distance between the two bands). We see the fragility of the topological phase $(m<1)$ in comparison with the rest of the cases $(m>1)$. The main plot corresponds to the eight-site CL, and the inset depicts the initial decay rate in the thermodynamic limit (see main text).}
	\label{fig:fidelity}
\end{figure}

\begin{figure}[h]
	\includegraphics[width=\columnwidth]{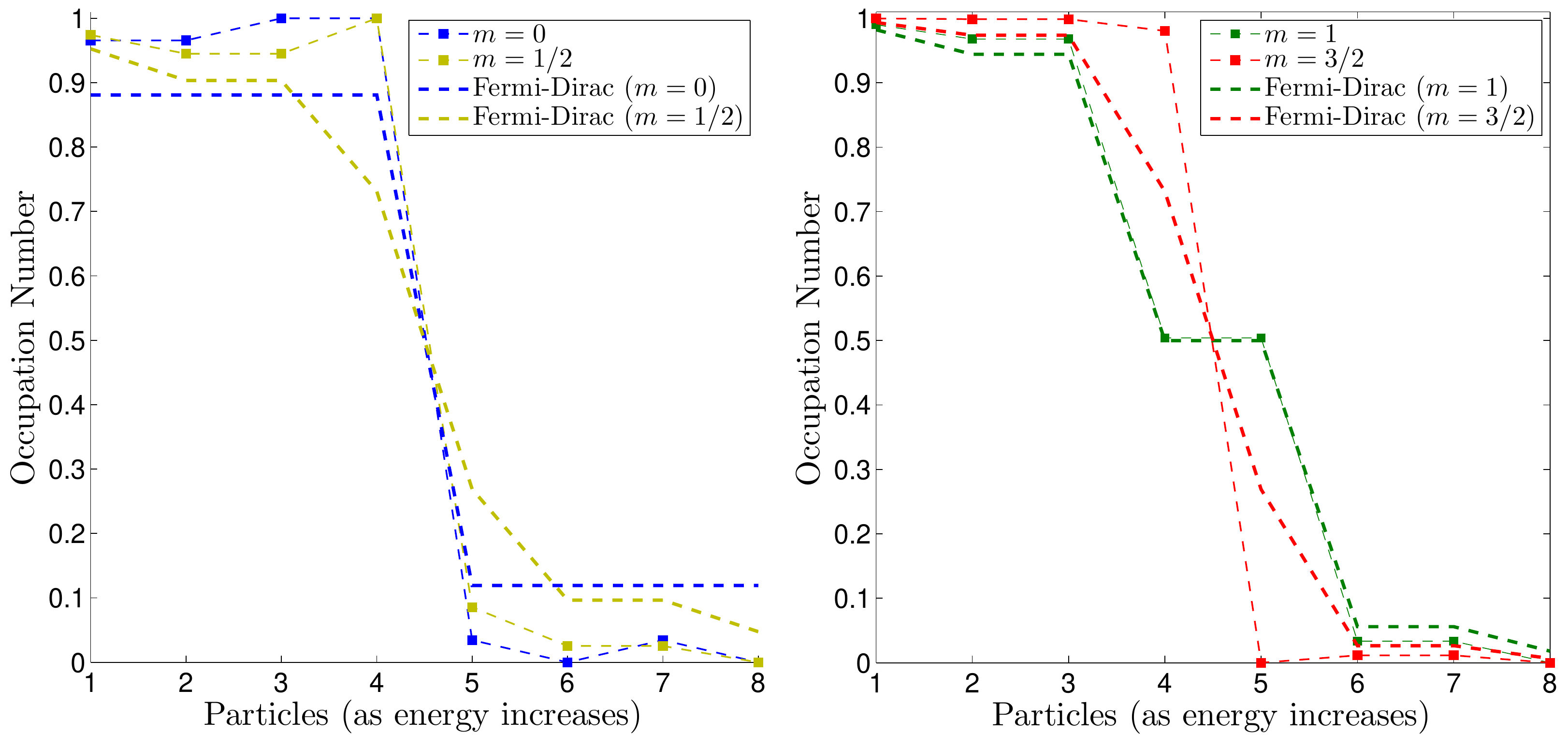}
	\caption{
Asymptotic occupation of the fermions for different values of $m$. The temperature is set to $T=1$ (in units of $K$, Eq. (2)). On the left side we have represented some examples which present topological order, on the right side the system is out of the topological phase. Note that on the left side the point $4$ is fixed whereas the point $5$ is fixed on the right side. For the sake of comparison we have depicted also the Fermi-Dirac distribution which corresponds to the thermal state.}
	\label{fig:occupation_numbers}
\end{figure}

\begin{figure}[h]
	\includegraphics[width=\columnwidth]{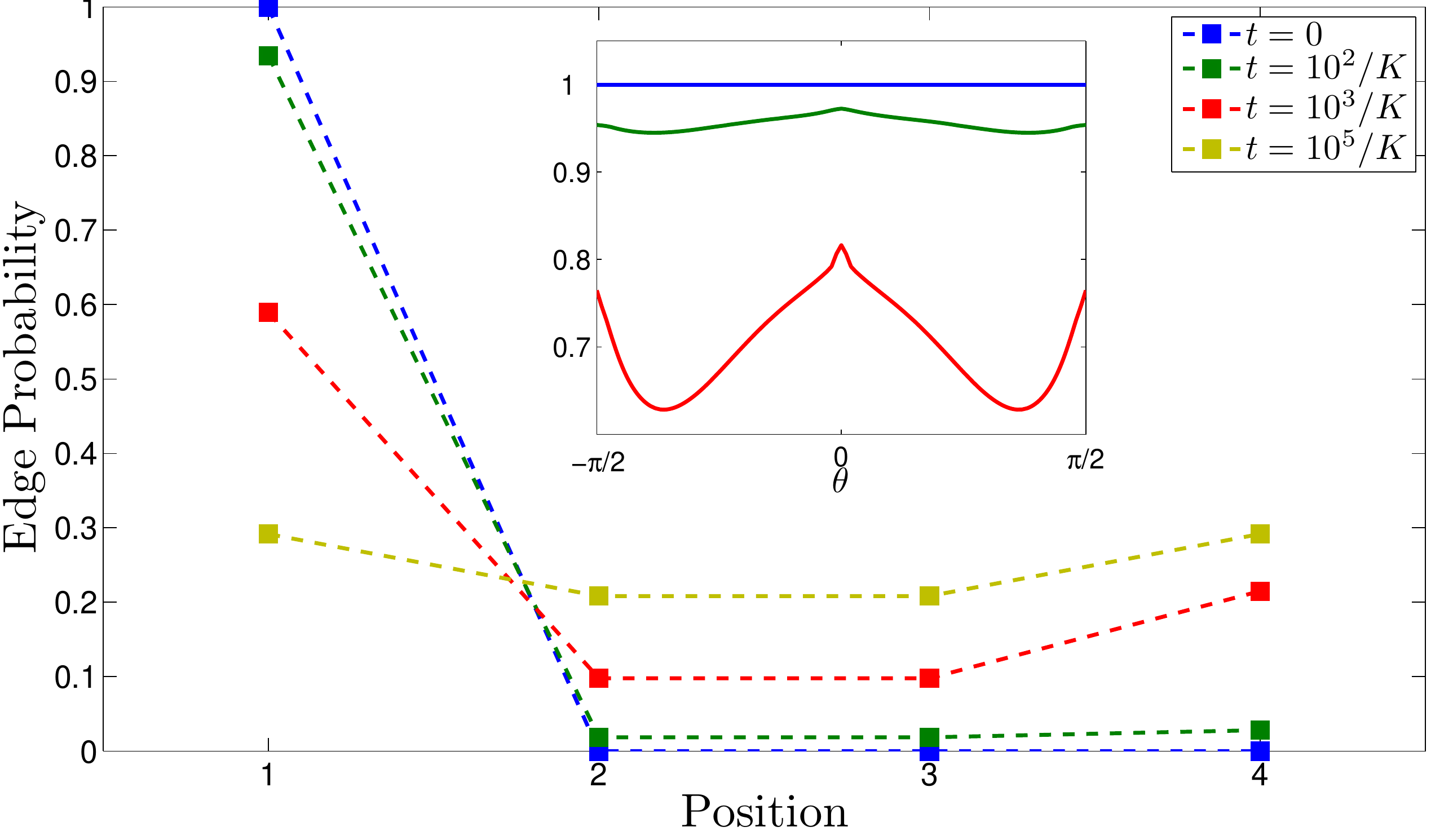}
	\caption{Instability of the topological end states against thermal noise in a eight-site CL (the six-site CL provides the same results). Here $\theta=\pi/2$ and the bath temperature is $T=0.04$ (in units of $K$, Eq. (2)). That is 1\% of the CL gap. The $x-$axis represent the longitudinal position on the CL. Thus, at $t=0$ all fermions are localized in left end, and as time increases they tend to delocalize along the whole chain. The inset plot depicts the probability that the fermions are at the end for several values of the magnetic flux $\theta$. The decay rate of the end state is smoother at $\theta=0$ and $\theta=\pi/2$ as the distance between bands increases for these values. Therefore, it is clear that the topological end states are unstable for any value of the magnetic flux.  Note that for the decay of a right-end state the result is symmetric because of the chiral symmetry of the interaction \eqref{Hint}.
}
\label{fig:edge_states}
\end{figure}

A common belief in TI theory is that the gap defining the topological phase is enough to protect
the system against interactions, disorder and even dynamical effects  \cite{hatch_11}, but the effects of dynamical thermal noise has not been addressed thus far:

\noindent i/  We have shown that it does not hold for finite temperature
effects: the TI order gets lost regardless of the gap size (see Fig. \ref{fig:fidelity} and Fig. \ref{fig:edge_states}).

\noindent ii/ Protected end states become unstable when the TI becomes a open quantum system couple to thermal baths. This is so even when
the interaction with the environment respects chiral symmetry, responsible for the robustness of end states in the TI (see Eq. \eqref{Hint}).
This is a highly non-trivial and novel effect. It implies that we have found a dynamical thermal
mechanism that is relevant to the description of the robustness of end states in TIs.
Prior to this work, the robustness of end states in TIs was solely judged on the basis
of the protecting symmetries in the isolated system.

\noindent iii/ We have observed that the existence of topological order strongly influences the system-bath interaction (Fig. \ref{fig:fidelity}). In particular, for our 1D model, the decoherence process remarkably depends whether the system is in a topological phase or not.

\noindent  iv/ Notably, the interaction with bosonic thermal baths does not lead this system to the thermal state. However the asymptotic state reached in the low temperature regime is close to it (see Fig. \ref{fig:occupation_numbers}).

%%%%%%%%%%%%%%%%%%%%%%%%%%
%%%%%%%%%%%%%%%%%%%%%%%%%%
%%%%%%%%%%%%%%%%%%%%%%%%%%
%%%%%%%%%%%%%%%%%%%%%%%%%%

Our fundamental result is the derivation of  the master equations \eqref{masterEq} and \eqref{decay_rate} for a TI under a bosonic thermal bath from which the thermal instability of the end states is derived along with other relevant consequences.

The total Hamiltonian of the problem considered reads as follows:

\begin{equation}
H=H_{\rm s}+H_{\rm bath}+H_{\rm int}.
\label{H}
\end{equation}
The first term, $H_{\rm s}$, is the Hamiltonian of the CL,
\begin{equation}
\label{creutz_hamiltonian}
\begin{split}
H_{\rm s}:=&-\sum_{n=1}^{N}\left[K\left(\ee^{-\ii\theta}a^{\dagger}_{n+1}a_n+\ee^{\ii\theta}b^{\dagger}_{n+1}b_n\right)+\right . \\
&\left. +K\left(b^{\dagger}_{n+1}a_n+a^{\dagger}_{n+1}b_n\right)+ M a^{\dagger}_{n}b_n+\text{h.c.}\right],
\end{split}
\end{equation}
where $a_n$ and $b_n$ are fermionic operators satisfying the anticommutation relations: $\{a_n,a^{\dagger}_{n'}\}=\delta_{n,n'}$, $\{b_n,b^{\dagger}_{n'}\}=\delta_{n,n'}$.
$K$ and $M$ are hopping amplitudes, while $\theta$ is the magnetic flux per plaquette in natural units. It is known that for $M <2K$ and open boundary conditions,
the system exhibits protected end states \cite{Creutz} that corresponds to a TI in 1D \cite{Berm1}.

The second term, $H_{\rm bath}$, is the free Hamiltonian of the local baths,
\begin{equation}
H_{\rm bath}:=\sum_{n,i}\epsilon^i_{n}A^{i\dagger}_{n}A^i_{n},
\end{equation}
where $A$ and $A^{\dagger}$ stand for the bath bosonic operators that satisfy the canonical commutation relations
$[A^i_n,A^{j\dagger}_{n'}]=\delta_{n,n'}\delta_{i,j}~,~[A^i_n,A^{j}_{n'}]=0$. Moreover, the index $n$ denotes the position of the local bath on the CL, and $i$ runs over the bath degrees of freedom.

Finally, the third term in \eqref{H}, $H_{\rm int}$, describes the interaction between the CL and the baths. In momentum space it reads (see Fig. \ref{fig:creutz_ladder})
\begin{equation}
H_{\rm int}:=\frac{1}{N}\sum_{i,k,k'}g^i_{kk'}(a^{\dagger}_k+b^{\dagger}_k)(a_{k'}+b_{k'})\otimes(A^{i{\dagger}}_{k'-k}+A^i_{k-k'}).
\label{Hint}
\end{equation}
The quantity $g^i_{k,k'}$ regulates the boson-fermion coupling, and it is chosen in such a way that chiral symmetry is preserved. More specifically, the chiral symmetry corresponds to a rotation of $\pi$ around the magnetic field axis \cite{Creutz}, which in momentum space swaps the fermion modes $a$ and $b$, and changes the sign of $k$ (see Fig. \ref{fig:creutz_ladder}). Thus, we shall assume $g^i_{k,k'}=g^i_{-k,-k'}$ so that the Hamiltonian \eqref{Hint} does not break chirality, which is indeed a very natural assumption for this coupling. 
Note that although the system contains free fermions in a gauge background field, its dynamics is highly nontrivial since the coupling with the bosonic bath involves three-body interactions (see Fig. \ref{fig:creutz_ladder}).

%%%%%%
\section{Master Equation for a 1D Topological Insulator}
%%%%%%

The evolution of system and bath is given by the Liouville-von Neumann equation, which in the interaction picture reads (unless otherwise stated, natural units $\hbar=k_{\rm B}=1$ are taken throughout the paper)
\begin{equation}
\frac{d\tilde{\rho}}{dt} =-{\rm i}[\tilde{H}_{\rm int},\tilde{\rho}],
\label{LE}
\end{equation}
where
\begin{eqnarray}
\tilde{H}_{\rm int}&=&{\rm e}^{{\rm i}(H_{\rm s}+H_{\rm bath})t}H_{\rm int}{\rm e}^{-{\rm i}(H_{\rm s}+H_{\rm bath})t}\nonumber
\\&=&\frac{1}{N}\sum_{i,k,k'}\sum^2_{\alpha,\beta=1}g^i_{kk'}(f^{\alpha\dagger}_kh^{\alpha\beta}_{k,k'}\text{e}^{{\rm i}\lambda^{kk'}_{\alpha\beta}t}f^{\beta}_{k'}) \nonumber \\
&&\otimes(\text{e}^{{\rm i}\epsilon^i_{k'-k}t}A^{i\dagger}_{k'-k}+\text{e}^{-{\rm i}\epsilon^i_{k-k'} t}A^i_{k-k'}).
\label{Hint2}
\end{eqnarray}
Here $f^{1}:=c$ and $f^{2}:=d$ are the operators which diagonalize the CL Hamiltonian; that is, $H_{\rm s}=\sum_{k\in{\rm BZ}}\lambda_1^k c^\dagger_k c_k+\lambda_2^k d^\dagger_k d_k$ with
\begin{equation}
\lambda^k_{1,2}:=2K\left[-\cos{k}\cos{\theta}\mp\sqrt{\sin^2{k}\sin^2{\theta}+(m+\cos{k})^2}\right]\\
\end{equation}
and $m:=M/2K$. Moreover $h^{\alpha\beta}_{k,k'}:=F^{\alpha}_kF^{\beta}_{k'}$, where
\begin{eqnarray}
F^{\alpha}_k &:=&\frac{x_k-(-1)^{\alpha}}{\sqrt{1+x^2_k}}, \\
x_k&:=&\frac{\sin{k}\sin{\theta}+\sqrt{\sin^2{k}\sin^2{\theta}+(m+\cos{k})^2}}{m+\cos{k}}.
\end{eqnarray}
Finally, $\lambda^{kk'}_{\alpha\beta}:=\lambda_\alpha^k-\lambda^{k'}_\beta$ are Bohr frequencies associated with the eigenvalues of the system Hamiltonian \eqref{creutz_hamiltonian} which represent the two energy bands.

%%%%%%%%%%%%%%%%%%%%%%%%%%
%%%%%%%%%%%%%%%%%%%%%%%%%%
%%%%%%%%%%%%%%%%%%%%%%%%%%
%%%%%%%%%%%%%%%%%%%%%%%%%%

By tracing out the bath's degrees of freedom from \eqref{LE}, we aim at writing a dynamical equation for the CL density matrix, $\rho_{\rm s}=\Tr_{\rm bath}(\rho)$. Under
 the natural assumptions  of the Born-Markov  coupling to the thermal bath  (\cite{Alicki,BrPe,Libro} and references therein), we arrive at the following master equation
 for the TI:
\begin{multline}\label{masterEq}
\frac{{\rm d}\rho_{\rm s}}{{\rm d}t} =-{\rm i[}H_s,\rho_s]+ \sum_{\substack{k,k'\\q,q'}}\sum^2_{\substack{\alpha,\beta\\ \gamma,\delta=1}}\Gamma^{kk'qq'}_{\alpha\beta\gamma\delta} (f^{\alpha{\dagger}}_kf^{\beta}_{k'}\rho_s(t)f^{\gamma{\dagger}}_qf^{\delta}_{q'} \\
-\frac{1}{2}\{f^{\gamma{\dagger}}_qf^{\delta}_{q'}f^{\alpha{\dagger}}_kf^{\beta}_{k'},\rho_s(t)\}),
\end{multline}
with
\begin{equation}
\begin{split}
\Gamma^{kk'qq'}_{\alpha\beta\gamma\delta}=\frac{1}{N^2}2\pi J(|\lambda^{qq'}_{\gamma\delta}|)[\Theta(\lambda^{qq'}_{\gamma\delta})+\bar{n}(|\lambda^{qq'}_{\gamma\delta}|)] \\
F^{\alpha}_{k}F^{\beta}_{k'}F^{\gamma}_{q}F^{\delta}_{q'}\delta_{\lambda_{\beta\alpha}^{k',k},\lambda_{\gamma\delta}^{q,q'}}\delta_{k'-k,q-q'}.
\end{split}
\label{decay_rate}
\end{equation}

These $\Gamma$'s are the decay rates induced by the dissipative dynamics in our system. Here, $\Theta(\omega)$ denotes the Heaviside step function and $\bar{n}(\omega)=\left[e^{\omega/T}-1\right]^{-1}$ is the number of bosons with frequency $\omega$ in each local bath; $T$ stands for the bath temperature. For the sake of simplicity we have assumed that $g^i_{k,k'}$ only depends on the difference between $k$ and $k'$, $g^i_{k,k'}\equiv g^i_\epsilon$, where the energy $\epsilon$ is related to $k-k'$ through the dispersion relation of the baths (note this is consistent with the chirality preserving condition $g^i_{k,k'}=g^i_{-k,-k'}$). In such a case, the so-called spectral density of the bath is formally written as $J(\omega):=\sum_i (g^i_\epsilon)^2\delta(\omega-\epsilon^i)$. For definiteness and as the CL could be realized in an optical lattice setup, we will consider a typical spectral density for a quantum optical 1D system, $J(\omega)=\alpha\omega \text{e}^{-\frac{\omega}{\omega_c}}$ where $\alpha$ is a parameter that regulates the interaction strength (typically $\alpha$ will be the fine-structure constant) and $\omega_c$ is a cutoff frequency. Furthermore, this  ``Ohmic'' spectral density is widely used in the modeling of condensed matter systems as well \cite{Weiss}.

Despite the apparent complicated structure of $\Gamma^{kk'qq'}_{\alpha\beta\gamma\delta}$ decay rates, the nonvanishing contributions are well understood and both numerical and analytical calculations can be carried out with precision.

%%%%%%%%%%%%%%%%%%%%%%%%%%
%%%%%%%%%%%%%%%%%%%%%%%%%%
%%%%%%%%%%%%%%%%%%%%%%%%%%
%%%%%%%%%%%%%%%%%%%%%%%%%%

%%%%%%%%%
\section{Non-Perturbative Thermal Dynamics}
%%%%%%%%%

In this section we analyze the out-of-equilibrium physics described by the master equation \eqref{masterEq} in the case of a finite size CL.
This allows us to retrieve nontrivial results about the stability of the topological order and whether it thermalizes, among other properties.
In order to study the stability of the system, we may chose different figures of merit. For instance, the fidelity ${\cal F}$ of the evolved mixed state of the system $\rho_{\rm s}(t)$ and the initial Fermi Sea (FS) for the lower band of the TI, represents a measure of how the system remains correlated to its initial state which exhibits a topological order:
\begin{equation}
{\cal F}\left[|{\rm FS}\rangle, \rho_{{\rm s}}(t)\right]:=\langle{\rm FS}|\text{e}^{t{\cal L}}\left(|{\rm FS}\rangle\langle{\rm FS}|\right)|{\rm FS}\rangle,
\label{fidelity}
\end{equation}
In fact, if this fidelity remains close to one then it is a strong indication that the topological order is preserved. Figure \ref{fig:fidelity} shows the behavior of the fidelity in a CL of size $N=8$. For some cases the fidelity may remain high, particularly for $m>1$, however in this case the CL is out of the topological phase \cite{Creutz,Berm1}. On the contrary, if $m<1$ the fidelity may be reduced up to $10\%$ of its initial value.

A complementary criterium for the topological order to persist is given by the evolution of the fermion occupation numbers. Most of fermions escaping from the lower band is a signature that the system no longer keeps the topological order. In figure \ref{fig:occupation_numbers} the occupation numbers are plotted for different values of $m$. The asymptotic occupation (i.e. occupation for sufficiently large times) is close to the Fermi-Dirac statistics, but they do not exactly fit each other.  
Finally, the existence of end states is a well-defined property that characterizes a TI. If they disappear after the TI is in contact with a thermal bath, we may unequivocally conclude that this type of topological order is lost.
As shown in Fig. \ref{fig:edge_states}, they are unstable and tend to delocalize along the chain in time. Note that if the bath temperature is small in comparison with the gap of the CL, the system takes a lot of time to delocalize. This fits with the very well-known argument that TI insulators are stable to perturbations if they present a large gap. However, they always delocalize under thermal noise after some sufficiently large period of time.

Let us see the implications of our thermal evolution analysis on the physical implementations of the
 Creutz ladder with fermionic atoms in an optical lattice  \cite{CL_OL_12}. For that purpose we need: two Zeeman sublevels attached to the fermion species $a_n$ and $b_n$ respectively; laser-assisted tunneling for the transversal and horizontal hopping $K$, and onsite Raman transitions for the vertical hopping $M$.
The thermal noise  can be a model for heating induced by lasers that create the optical trap (due to fluctuating intensity profiles), or any other type of bosonic thermal noise. We take experimental values for $m=0$ and $\theta=\pi/2$ as in \cite{1D_TI_OL_12}: $K\sim 3~\hbar$ kHz, gap $\Delta=12\hbar$ kHz, and bath temperature $T\sim56$ nK which are currently reachable.
 We obtain a lifetime for end states  $\tau\sim 67$ ms which is much larger than the typical time for the system dynamics $O(1/K)$. Hence, considering this type of thermal noise, measuring topologically ordered states could be possible within an optical lattice setup.

%%%%%
\section{Fidelity in the Thermodynamic Limit}
%%%%%%

The evolution of the state can be written up to second order in time as:
\begin{equation}
\rho_{{\rm s}}(t)=\text{e}^{t{\cal L}}\rho_{{\rm s}}(0)\simeq (1+t{\cal L}+\frac{t^2}{2}{\cal L}^2+...)\rho_{{\rm s}}(0).
\label{fidelity_perturbative}
\end{equation}
Using \eqref{fidelity},
and after several calculations with
the master equation \eqref{masterEq}, we obtain the following result
\begin{equation}
{\cal F}(|{\rm FS}\rangle\langle{\rm FS}|, \rho_{{\rm s}}(t))\simeq1-t\Gamma_{(1)}+\frac{t^2}{2}\left(\Gamma_{(1)}^2+\Gamma_{(2)}\right),
\end{equation}
where
\begin{equation}
\begin{split}
\Gamma_{(1)}:=\sum_{k,q}\Gamma^{kqqk}_{2112},\quad
\Gamma_{(2)}:=\sum_{k,q}\Gamma^{kqqk}_{2112}\Gamma^{qkkq}_{1221}.\\
\end{split}
\end{equation}

At short times, two rates $\Gamma_{(1)}$ and $\Gamma_{(2)}$ will determine how fast the fidelity of the Fermi Sea is lost during its evolution. The initial linear behavior for the lost of fidelity given by $\Gamma_{(1)}$ is patent in Fig. \ref{fig:fidelity} for $N=8$ as well. Direct processes exciting electrons from one band to the other are dominant in the dissipative evolution as we might expect. Furthermore, in the inset to Fig. \ref{fig:fidelity},  we can see that the initial decay of the Fermi Sea's fidelity strongly depends on whether or not the system is in a topological phase $m<1$. More explicitly, the decay of fidelity increases as we approach the topological crossover point $m=1$, and then the decay decreases significantly for $m>1$ -- out of the topologically ordered regime. This perturbative analysis for the thermodynamic limit is in total agreement with the exact results for $N=8$ as shown in Fig. \ref{fig:fidelity}, and with size $N=6$ (not shown).

%%%%%%
\section{Conclusions} 
%%%%%%
We have derived a master equation describing the dynamical thermal effects of bosonic baths coupled to a one-dimensional TI. As this
coupled fermionic-bosonic system is not exactly solvable, our formalism is useful to address relevant thermal effects of TIs in 1D.
Let us emphasize that our approach to studying thermal effects on TIs is beyond the standard formalism of assuming
that the system is in a thermal state at a certain temperature $T$. On the contrary, our purpose is to study the out-of-equilibrium
dynamics of a TI coupled to a thermal bath. It is this bath which has a well-defined temperature $T$ and disturbs the TI.
Very remarkably, the interaction with the thermal environment (fermion-boson) respects chiral symmetry.
This symmetry is responsible for the protection (robustness) of the end states in the topological insulator
when it is isolated from the environment. Therefore, our mechanism makes end states unstable while preserving
their protecting symmetry.
In addition, we observed that the dissipative dynamics distinguishes whether the system is in a topological phase or not.
We have also shown that thermal noise
delocalizes the topological end states into the bulk bands of the TI for sufficiently large times and regardless of the gap size. While this is compatible with the existence
of TIs in experiments, this thermal instability will play an important role in detailed control manipulations needed for quantum computation.

\begin{acknowledgments}

We thank the Spanish MICINN grant FIS2009-10061,
CAM research consortium QUITEMAD S2009-ESP-1594, European Commission
PICC: FP7 2007-2013, Grant No.~249958, UCM-BS grant GICC-910758.

\end{acknowledgments}


\begin{thebibliography}{99}
\bibitem{comment1} It realizes the symmetry class AIII (chiral unitary)  \cite{CL_OL_12}.
\end{thebibliography}
\end{document}